\documentclass[12pt]{article}
\usepackage{amsfonts,amssymb}

\renewcommand{\baselinestretch}{1.1}
\newcommand{\CP}[1]{\mathbb{P}^{#1}}
\newcommand{\C}[1]{\mathbb{C}^{#1}}

\newcommand{\R}[1]{\mathbb{R}^{#1}}

\newcommand{\WP}[1]{{\mathbb{WP}({#1})}}
\def\z{\zeta}
\def\Z{\mathbb{Z}}

\def\Y{\mathbf{Y}}
\def\S{\mathbf{S}}
\def\T11{{T}^{1,1}}
\def\be{\begin{equation}}
\def\ee{\end{equation}}
\def\bear{\begin{eqnarray}}
\def\eear{\end{eqnarray}}
\def\nn{\nonumber}
\def\ie{{\it i.e.\/}}

\def\bz{\mathbf{z}}
\def\bV{\mathbf{V}}

\def\bX{\mathbf{X}}
\def\cP{\mathcal{P}}

\def\dim{\mathrm{dim}}
\def\Vol{\mathrm{Vol}}
\def\vol{\mathrm{vol}}
\def\bg{\mathbf{g}}
\def\bh{\mathbf{h}}
\def\bw{\mathbf{w}}
\def\bR{\mathbf{R}}
\def\om{\omega}
\def\wdg{\wedge}
\def\st{\star}
\def\mO{\mathcal{O}}
\def\mcO{\mO}
\def\jj{\jmath}
\def\mcL{\mathcal{L}}
\def\del{\partial}

\def\Tr{\mathrm{Tr}}
\def\Id{\mathrm{Id}}
\def\ord{\mathrm{ord}}
\def\bi{\bibitem}
\usepackage{graphicx}

\begin{document}

\begin{titlepage}

\begin{flushright}
hep-th/0108020\\
PUPT-2002
\end{flushright}
\vfil

\begin{center}
{\huge The Volume of some Non-spherical Horizons }\\
\vspace{3mm}
{\huge and the AdS/CFT Correspondence}

\end{center}

\vfil
\begin{center}
{\large Aaron Bergman
and Christopher P. Herzog}\\
\vspace{1mm}
Joseph Henry Laboratories,  \\
Princeton University,\\
Princeton, New Jersey 08544, USA\\
{\tt abergman,cpherzog@princeton.edu}\\
\vspace{3mm}
\end{center}

\vfil

\begin{center}
{\large Abstract}
\end{center}

\noindent
We calculate the volumes of a large class of Einstein manifolds,
namely Sasaki-Einstein manifolds which are the bases
of Ricci-flat affine cones described by polynomial embedding relations
in $\C{n}$.  These volumes are important because they allow us to 
extend and test the AdS/CFT correspondence.  We use these volumes to
extend the central
charge calculation of Gubser (1998) to the generalized conifolds of
Gubser, Shatashvili, and Nekrasov (1999).  These volumes also allow one
to quantize precisely the D-brane flux of the AdS supergravity solution.
We end by demonstrating a relationship between the volumes of these 
Einstein spaces and the number of holomorphic polynomials (which 
correspond to chiral primary operators in
the field theory dual) on the corresponding affine cone.
\vfil
\begin{flushleft}
August 2001
\end{flushleft}
\vfil
\end{titlepage}
\newpage
\renewcommand{\baselinestretch}{1.1}  


\section{Introduction}

The AdS/CFT correspondence \cite{jthroat, gkp, EW} is motivated by comparing 
a stack of elementary branes with the metric it produces (for reviews, see, for
example, \cite{magoo, Krev}).  In order to break some of the supersymmetry,
we may place the stack at a conical singularity 
\cite{LNV, KS, KW, Kehag, MoPle}.
Consider, for instance, a stack of D3-branes placed at the apex of a
Ricci-flat six dimensional cone $\Y_6$ whose base is a 
five dimensional Einstein manifold $\bX_5$.  
Comparing the metric with the D-brane description leads one to conjecture
that type IIB string theory on $AdS_5 \times \bX_5$ is dual 
to the low-energy limit of the worldvolume theory on the D3-branes at
the singularity.  One may also consider a stack of M2-branes placed at the
apex of a Ricci-flat eight dimensional cone $\Y_8$ whose base is a
seven dimensional Einstein manifold
$\bX_7$.  There is a similar conjectured correspondence between M-theory
on an $AdS_4 \times \bX_7$ background and the low-energy limit of
the worldvolume
theory of the M2-branes at the singularity.  

In the simplest example of AdS/CFT correspondence, the D3-branes form a stack
in ten dimensional space.  The Einstein manifold is $\S^5$,
and the theory preserves ${\cal N}=4$ supersymmetry (SUSY).  
Subsequently, it was realized that 
certain orbifolds of $\S^5$, 
$\S^5 / \Gamma$ where $\Gamma$ is a discrete
subgroup of $SU(2)$, preserve ${\cal N}=2$ supersymmetry \cite{LNV, KS}.
The manifold $\bX_5 = \T11$ was examined in \cite{KW} and gives
$\mathcal{N} = 1$ supersymmetry. (This example and a number of
others were also examined in \cite{MoPle}.)
Unfortunately, $S^5$, $T^{1,1}$, and their orbifolds just about
exhaust the mathematical literature of five dimensional Einstein spaces for
which explicit metrics are known.  In these cases, it 
was explicit knowledge of the metric which allowed for many tests of the 
AdS/CFT correspondence.

Because of extensive work on compactifying eleven dimensional supergravity
to four dimensions in the eighties \cite{classification} and because
seven dimensions allow for a larger variety of homogeneous spaces than do five,
the metric situation is slightly better for the AdS/CFT correspondence 
with M2-branes.  Metrics are known for the seven dimensional Einstein
spaces $Q^{1,1,1}$, $M^{1,1,1}$, $N^{0,1,0}$, $V_{5,2}$, and of course
$\S^7$ among a few others as well.  Moreover, many tests of the AdS/CFT 
correspondence have been carried out using these metrics. 

The set of spaces described above is somewhat limited, and
it would
be useful to broaden the number of examples for which 
concrete calculations can be carried out which test and strengthen these
correspondences.
Encouragingly, many papers have appeared in the literature recently 
(for example \cite{GKP, ICV}) where explicit knowledge of the metric
was not needed, and valuable information about the corresponding gauge theory
duals was extracted from the geometry 
in other ways.  Continuing this trend, we show how to calculate the volume
of a large class of Einstein spaces without knowing an explicit metric
on them.
We then use these volumes to extend and test the AdS/CFT correspondence.

The large class of Einstein spaces we are concerned with are bases of certain
Ricci-flat affine cones.  Consider a 
weighted homogeneous polynomial in $\C{n+1}$, by which we mean a
polynomial $F(\bz)$ which satisfies
\[
F(\lambda^{w_0} z_0, \lambda^{w_1} z_1 , \ldots , \lambda^{w_n} z_n) =
\lambda^d F(z_0, z_1, \ldots, z_n) \ ,
\] 
where $\lambda \in \C{*}$ and $w_i \in \Z^+$, and the 
degree $d$ is a positive
integer.  
These spaces are cones because of the scaling with respect to $\lambda$.
Thus, we can write the metric on these spaces as
\[
ds_\Y^2 = dr^2 + r^2 ds_\bX^2 \ .
\]
The tensor $ds_\bX^2$ gives a metric on the intersection of this
cone with the unit sphere in $\C{n+1}$. Our formula gives the
volume of the intersection manifold endowed with this metric.

These volumes are important for at least two reasons.  
First, they allow us to determine the central charge of the dual
gauge theory. It was
conjectured in \cite{KW} that the gauge theory corresponding to an 
$AdS_5 \times T^{1,1}$
background can be obtained as the IR fixed point of a
renormalization group (RG) flow from the $\S^5 / \Z_2$ orbifold theory.  
Gubser realized \cite{Gubser} that this flow
had calculable consequences for the central charge of the two theories,
namely, 
\be
\frac{c_{IR}}{c_{UV}} = \frac{1/\Vol(T^{1,1})}{1/\Vol(\S^5 / \Z_2)} \ .
\ee
Later, extending this work, 
the authors of \cite{GNS} conjectured that the ${\cal N}=2$ orbifolds
$\S^5 / \Gamma$ flow to certain generalized conifolds
$\Y_\Gamma$.\footnote{The cones for the $A_k$ groups were also
derived in \cite{EL}.} However,
as the volumes of these generalized conifolds were unknown, the same
central charge calculation could not be repeated.  
Our volume formula applies to these generalized
conifolds. We show that the ratio of the central
charges for these generalized conifolds is exactly as predicted by the AdS/CFT 
correspondence.

Second, the volumes allow us to quantize precisely the flux in the supergravity
solutions.  It is known that for a stack of D3-branes placed at the conical singularity of $\Y_6$, the supergravity solution is
\be
ds^2 = h(r)^{-1/2} (-dt^2 + dx_1^2 + dx_2^2 + dx_3^2) + 
h(r)^{1/2} (dr^2 + r^2 ds_{X_5}^2) \ ,
\ee
\be
h(r) = 1+ \frac{4\pi^4 g_s N \alpha'^2}{\Vol(\bX_5) r^4} \ ,
\ee
\be
F_5 = {\cal F}_5 + \star {\cal F}_5 \ , \; \; \; 
{\cal F}_5 = 16\pi \alpha'^2 N \frac{\Vol(\S^5)}{\Vol(\bX_5)} \vol(\bX_5) \ ,
\ee
where all the other field strengths vanish and $N$ is the number of D3-branes.
With this notation, $\vol$ is the volume differential form.  Thus
\[
\int_{\bX_5} \vol(\bX_5) = \Vol(\bX_5) \ .
\]
Presumably, other uses for these volumes can be found.

We finish with an intriguing relationship between the volume of our 
Einstein spaces and the number of holomorphic monomials of a given 
total degree $L$ on the corresponding affine cone.  For affine cones
of a given complex dimension $n$ and a given index which
is the sum of the weights $w_i$ minus the degree $d$, the volume and number
of these holomorphic monomials are directly proportional to each
other.  Although the relationship
is generally true, we have only directly shown``why'' it is true in the
smooth case. The proof should generalize, but we have not done so here.
The relationship is intriguing because these holomorphic monomials
play an important role in AdS/CFT correspondence.  They are chiral primary
operators, and supersymmetry protects their dimension allowing for direct
comparison between the gauge theory and the supergravity dual.

This paper is organized as follows. We begin with the
derivation of our general volume formula. In section 3, we
apply this formula to the case of the generalized conifolds of
\cite{GNS} and show that the AdS/CFT correspondence gives the
correct result for the ratio of central charges. Finally,
we discuss the relation between the
volume of these spaces and the number of holomorphic monomials
and the mathematical reason why this conjecture is true.

\section{The Volume of a Large Class of Einstein Manifolds}


In order to understand the volume computation, it will be useful to keep
the following picture in mind. We recall from above that there
is a natural $\C{*}$ action on our cones.
If we quotient out by the $\R{+}$ portion of $\C{*}$, we
get a manifold that is termed the base of the cone and which we denote
$\bX$. If we further quotient out by the remaining $U(1)$ part
of $\C{*}$, we obtain the weighted projective variety defined by
$F(\bz)=0$ in the weighted projective space $\WP{\bw}$. We will
call this manifold $\bV$.

Because there exists a Calabi-Yau metric on
our original cone, there is an
Sasaki-Einstein metric on $\bX$ and an K\"ahler-Einstein metric
on $\bV$. In addition, $\bX$ is a $U(1)$ fibration over $\bV$. We
will use the K\"ahler-Einstein condition to determine the
volume of $\bV$, and we will use the 
metric on $\bX$ to determine
the length of the fiber and thus the total volume of $\bX$.
Unfortunately, $\bV$ may not in general be a manifold. In fact,
it will be an orbifold.\footnote{An orbifold is a space which
looks locally like the quotient of $\R{n}$ by a finite group. In
other words, over each open set in a sufficiently fine open
cover of the orbifold, there exists a covering subset of $\R{n}$
and a finite group such that
we identify the set in the orbifold with the quotient of the covering
set by the action of the
group. The open cover of the orbifold and the collection of
covering sets and groups is termed the local uniformizing system
of the orbifold. A V-bundle generalizes the concept of a fiber
bundle to orbifolds and consists of a fiber bundle over each
covering set and certain equivariant gluing relations.
See the references for more details.}
In order to accommodate this difficulty,
we take a brief excursion to introduce some aspects of weighted
projective spaces and hypersurfaces in them.

\subsection{Weighted Projective Spaces}

A weighted projective space is defined in analogy to
ordinary projective space: instead of a uniform weighting,
the $\C{*}$ action on
$\C{n+1}$ is weighted by a vector of weights $\bw$ as above. We denote this
space $\WP{\bw} \equiv \WP{w_0,\ldots,w_n}$.
Let us also write $|\bw| = \sum w_i$ and $w = \prod w_i$. Any
polynomial that is homogeneous under the weighted action defines
a weighted projective variety. Under certain conditions, one can
treat weighted projective varieties similarly to
ordinary projective varieties. For a review, see, for
example, \cite{Dol,Fle}. In these references (see also
\cite{BG3,BG4}), one finds conditions for the hypersurface
to be well-formed and quasismooth. The former requires that 
\begin{eqnarray}
\label{wf1}
&\mathrm{gcd}(w_0,\ldots,\hat{w_i},\ldots,\hat{w_j},\ldots,w_n)
\mid d& \\
\label{wf2}
&\mathrm{gcd}(w_0,\ldots,\hat{w_i},\ldots,w_n) = 1&
\end{eqnarray}
where a hat means to omit an element. This condition ensures
that the singularities of the hypersurface are of complex codimension 2
or greater. Note that any weighted projective space is always
isomorphic to one for which (\ref{wf2}) holds \cite{Dol,Fle}.
The conditions on the weights for quasismoothness
are technical and not particularly elucidating, so we will refer
the reader to the references for a full discussion. However,
one can formulate the condition of quasismoothness in a manner
that conforms to our physical expectation. Basically, a
hypersurface is quasismooth if the affine cone over it is smooth
at all points except the vertex \cite{Fle}. In other words, this
condition is just the statement that the only nonsmooth point of
the cone on which the D-branes live is its tip where the D-branes
are placed.
This is altogether reasonable. All the spaces that we will deal with
in this paper satisfy these conditions.

The main consequence of these conditions is that the standard
adjunction formula from algebraic geometry holds for these
hypersurfaces:
\be
\label{adj}
\mO(K^{-1}) = \mO(|\bw| - d)
\ee
where $K^{-1}$ is the dual of the canonical sheaf, and $d$ is the
degree of the weighted homogeneous polynomial. The quantity
$|\bw| - d$ coincides with the index of the orbifold. It is a theorem of
\cite{BG2} that there exists a line V-bundle, $H$, such that
$H^{|\bw|-d} = K^{-1}$. We will call $H$ the hyperplane V-bundle.
A degree $d$ hypersurface is the zero locus of the $d$th power

By working on the local uniformizing system of the orbifold, we
can also define a connection on a given V-bundle. This allows us to 
define characteristic classes by means of the Chern-Weil
homomorphism (see, for example, \cite{Nak}) which defines
characteristic classes in terms of symmetric polynomials in the
curvature of a connection on the given bundle.
Because we are in
the orbifold category, however, these will be defined over the
rationals rather than over the integers. There is a
definition of orbifold cohomology due to Haefliger \cite{Hae},
but we will not refer to it here. Orbifold and ordinary cohomology
are, in fact, isomorphic over the rationals. 

\subsection{The Geometry of Calabi-Yau Cones}

We now review the features of Calabi-Yau cones that we will use in the
sequel. Because of the $\R{+}$ action on the cones, we can write
the Calabi-Yau metric as
$$
ds^2 = dr^2 + r^2\ g_{ab}\ dx^a\ dx^b.
$$
The base of the cone, $\bX$, is simply the intersection of the
cone with the unit sphere in $\C{n+1}$. The tensor $\bg = g_{ab}
dx^a dx^b$ defines a metric on $\bX$. This is one definition of
a Sasaki manifold. Sasaki manifolds have many special
properties and are reviewed in \cite{BG3}. Because our
cone is Calabi-Yau and, hence, Ricci-flat, it is easy to see
that the metric $\bg$ must be Einstein with scalar curvature
$s = 2(n-1)(2n-1)$ where $n-1$ is the complex dimension of $\bV$ and $2n-1$
is the real dimension of $\bX$. Sasaki-Einstein
manifolds are reviewed in \cite{BG3}. Recall that an Einstein
manifold satisfies the relation
\be
\label{erel}
\bR = \frac{s}{\dim_\R{}} \bg
\ee
where $\bR$ is the Ricci tensor on the manifold.

The main result we will use from the above papers is that there is
a canonical foliation of $\bX$ by circles and that the space of
leaves, $\bV$, is an $n-1$ complex dimensional complex orbifold with a
K\"ahler-Einstein metric, $\bh$, of scalar curvature $4n(n-1)$.

In fact, we have more structure here. Our cones are cut out by
weighted homogeneous polynomials in complex space. This situation
has been extensively studied in \cite{BG3,BG4}. Here, the space
of leaves, $\bV$, is exactly the weighted projective variety cut
out by the polynomial in $\WP{\bw}$. The foliation is just the
$U(1)$ action on $\bX$ inherited from the original $\C{*}$
action on the cone. The inversion theorem (theorem 2.8)
of \cite{BG2} tells us
that $\bX$ is a $U(1)$ V-bundle over $\bV$ with the Sasaki-Einstein metric
\be
\label{semet}
\bg = \pi^* \bh + \eta \otimes \eta
\ee
where $\pi^*$ denotes the pullback from the base to the
fibration, and $\eta$ is a connection 1-form on the fibration
with curvature $d\eta = 2\pi^* \omega$ where $\omega$ is the
K\"ahler form of the K\"ahler-Einstein metric on the base. For
more information on how these definitions generalize to the
orbifold category, see, for example, any of the above papers or
the original papers \cite{Sat1,Sat2,Bai1,Bai2}.

\subsection{Computing the Volume}

\subsubsection{The Volume of $\bV$}

Before determining the volume of $\bX$, we will first
determine the volume of $\bV$. The
Einstein relation will be the key feature that allows us to
determine the volume without an explicit knowledge of the
metric.

First, we recall a few definitions. As we are working on a
complex manifold, we can write our metric in complex coordinates
$h_{a\bar{b}}$. Thus, the K\"ahler form is given by
$$
\omega = i\ h_{a\bar{b}}\ dz^a \wdg d\bar{z}^{\bar{b}}.
$$
Also, we denote the volume form on the manifold by $\st 1$. A
simple calculation gives that $\st 1 = \om^{n-1} / (n-1)!$. The first Chern
class of a manifold, denoted by $c_1(\bV)$, is given in local
coordinates by:
\be
\label{chdef}
c_{1}(\bV) = i\frac{R_{a\bar{b}}}{2\pi} \ dz^{a} \wedge
d\bar{z}^{\bar{b}}
\ee
where $R_{a\bar{b}}$ is the Ricci tensor in complex coordinates.

We now will use the Einstein relation to relate $c_1(\bV)$ to
$\om$. As the scalar curvature of $\bV$ is $4n(n-1)$, the Einstein 
relation takes the form
$$
\bR = 2n \bh
$$
which implies that
\be
\label{chrel}
\om = \frac{\pi}{n} c_1(\bV).
\ee
So, finally, we have
\be
\label{volalm}
\Vol(\bV) = \int_\bV \st 1 =
\frac{\pi^{n-1}}{(n-1)! n^{n-1}} \int_\bV c_{1}(\bV)^{n-1}.
\ee

It finally remains to determine the integral of the Chern
classes. This requires a bit more investment than the rest of
the calculation, so we will consign the derivation to an
appendix. The end result is that 
\be
\label{chint}
\int_\bV c_{1}(\bV)^{n-1} = \frac{d}{w} (|\bw|-d)^{n-1}.
\ee
Combining this with (\ref{volalm}), we can now calculate the volume of $\bV$:
\be
\label{vol}
\Vol(\bV) = \frac{d}{w (n-1)!} 
\left(
\frac {\pi(|\bw|-d)}{n}
\right)^{n-1} \ .
\ee

\subsubsection{The Volume of $\bX$}

The last step that remains is to determine the length of the
fiber. While we will write as if the base, $\bV$, were a manifold,
all the following steps can be justified by working in the
uniformizing charts of the orbifold. We recall from above that
we have a partially explicit form of the metric on $\bX$, (\ref{semet}).
Let $\phi$ be a coordinate along the fiber. Then, we can write
$\eta = d\phi - \sigma$ where $d\sigma = 2\om = (2\pi/n)
c_1(\bV)$. An elementary
fact is that iff $\phi \in [0,2\pi]$ then
\be
\label{lendef}
d\sigma = 2\pi c_1
\ee
where $c_1$ is the first Chern class of the circle fibration.

In order to determine $c_1$ of the fibration, we recall our
picture of the cone with a $\C{*}$ action. The weighted
projective space that $\bV$ lives in is just the quotient of the
ambient space that the cone lives in by the same action. Thus,
we see that the circle fibration must have the same first Chern
class as the dual of the hyperplane V-bundle.\footnote{It is worthwhile to
note that one can make other choices for this Chern class. In the
smooth case, the allowed values can be determined from the Gysin 
sequence of the fibration. In the five dimensional case, as all
four dimensional K\"ahler-Einstein manifolds are known, one
obtains a complete classification of five dimensional regular
Sasaki-Einstein manifolds. For more details, see
\cite{FrK1,FrK2}. The role of these spaces in the AdS/CFT
correspondence is treated in some detail in \cite{MoPle}.}
From the adjunction formula (\ref{adj}),
we have $c_1 = -c_1(H) = -c_1(\bV)/(|\bw|-d)$. Therefore, it is
clear that the length of the fiber is not $2\pi$. In order to
remedy this, we will rescale the coordinate along the fiber to
make the relation (\ref{lendef}) hold. In particular, let
$\theta = -\phi n/(|\bw|-d)$. Then, the metric (\ref{semet})
takes the form 
$$
\bg = \pi^* \bh + \left(\frac{|\bw|-d}{n}\right)^2
(d\theta - \sigma')^2
$$
Now, we have
$$
d\sigma' = \frac{-n}{|\bw|-d} d\sigma = \frac{-2\pi}{|\bw|-d} c_1(\bV)
= 2\pi c_1
$$
which means that the coordinate length of the fiber is now $2\pi$.
We can easily do the integration and determine the geodesic
length of the fiber to be $2\pi\frac{|\bw| -d}{n}$. Combining
this with (\ref{vol}), we obtain a general formula for the
volume of the base of an affine cone over a weighted projective
variety:
\be
\label{voltot}
\Vol(\bX) = 2\pi\frac{|\bw|-d}{n} \Vol(\bV) =
\frac{2d}{w(n-1)!}\left(\frac{\pi(|\bw|-d)}{n}\right)^{n}.
\ee
This formula should generalize in a straightforward manner to
complete intersections, but we have not done so here. The
adjunction formula does hold for complete intersections in
weighted projective space \cite{Dol}.

\subsection{Checks of the Volume Formula}

It is interesting to check that we get the expected volumes from (\ref{voltot})
in some simple cases.

Consider first the hypersurface defined by
\[
F = \sum_{i=0}^n z_i = 0 \ .
\]
A moment's reflection should convince the reader that $F$ cuts out 
a copy of $\C{n}$ inside $\C{n+1}$. As a result, the corresponding
Einstein manifold is a $2n-1$ dimensional sphere.  The volume formula in
this case gives
\[
\Vol(\S^{2n-1}) = \frac{2 \pi^n}{(n-1)!} 
\]
which is indeed the correct answer.

Some less trivial examples to consider are the
Stenzel manifolds \cite{Stenzel}
\[
F = \sum_{i=0}^n z_i^2 = 0 \ .
\]
In the case $n=2$, note that the Ricci flat metric on 
$x^2 + y^2 + z^2 = \epsilon^2$ where $\epsilon \in \R{}$ is the
Eguchi-Hanson metric.  The case $\epsilon=0$ is well known to
correspond to $\R{4}/\Z_2$.  Indeed,
from (\ref{voltot}), we find that
\[
\Vol(\S^3 / \Z_2) = \pi^2 
\]
as it should be.

The next Stenzel manifold, $n=3$, is the well known conifold.
As a result, $\bX^5 = T^{1,1}$.  From the explicit metric
on $T^{1,1}$, or from (\ref{voltot}), one may calculate
that 
\[
\Vol(T^{1,1}) = \frac{16 \pi^3}{27} \ .
\]

For $n=4$, $\bX_5 = V_{5,2}$, one of the Stiefel manifolds.
Again, explicit metrics on $V_{5,2}$ are known \cite{CGLP, Stenzel}.  
One may
calculate from the metric, as we have done in the appendix,
or from (\ref{voltot}) that
\[
\Vol(V_{5,2}) = \frac{27 \pi^4}{128} \ .
\]

So far, we have only checked the volume formula for the smooth cases, \ie, 
the cases where $F(\bz)$ is a homogeneous polynomial.  There is one set
of simple singular example that we can also check.
Consider the weighted polynomials $F(x,y,z)$ in the complex
variables $x$, $y$, and $z$ of total degree $h$.  
Let these variables transform with the weights
$\alpha$, $\beta$, and $h/2$ given in (\ref{weights}). While the
weights here do not satisfy (\ref{wf2}), as noted above, the
spaces are isomorphic to ones for which the condition holds. The polynomials
describe ALE spaces and have an ADE classification (see for example
\cite{GNS,Dol}).  The Einstein manifolds at the bases of these cones are 
well known to be orbifolds of $\S^3$.  
The orbifold groups $\Gamma$ are discrete subgroups of SU(2), and  
each group $\Gamma$ has $2 \alpha \beta$ elements.  As a result
\be
\Vol(\S^3 / \Gamma) = \frac{\pi^2}{\alpha \beta} \ .
\label{alevol}
\ee
Let us compare this result with the volume formula (\ref{voltot}).
Note from (\ref{weights}) 
that the index $ |\bw| - d =  \alpha + \beta + h/2 - h$ 
is in every case equal to one.  
The products of the weights in (\ref{voltot}), $\alpha \beta h/2$,
cancels with the total degree $h$ to leave the required factor
of $\alpha \beta$ in the denominator of (\ref{alevol}).  We get
precisely the correct answer.

\section{A Central Charge Calculation}

We now wish to use our formula to check the ratio of central
charges for the RG flows of \cite{GNS}. We first compute the answer on
the field theory side and then compare with the prediction of
our volume formula.

\subsection{A Gauge Theory Perspective}

\begin{figure}
\includegraphics[width=\textwidth]{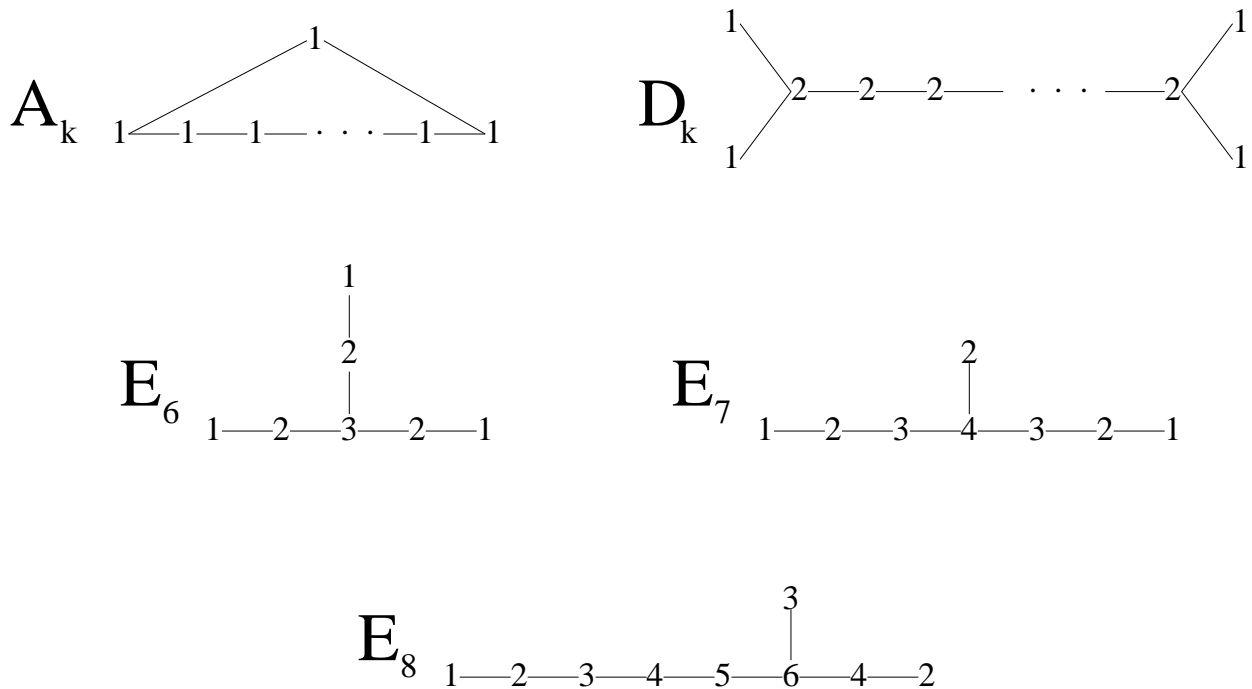}
\caption{The extended Dynkin diagrams of ADE type, including the
         indices $n_i$ of each vertex.}
\label{fig1}
\end{figure}

Let us begin by reviewing the gauge theory on the worldvolume
of a collection of 
$N$ D-branes placed at the orbifold singularity of $\C{2}/\Gamma$, where $\Gamma$ is a 
discrete subgroup of $SU(2)$ of ADE type.\footnote{Much of this
discussion is drawn from \cite{GNS}.}  The field theory has ${\cal N}=2$ 
supersymmetry.  Its gauge group is the product
\[
G = \times_{i=0}^{r}U(N_i)
\]
where $i$ runs through the set of vertices of the extended Dynkin diagram of the 
corresponding ADE type (see figure \ref{fig1}) \cite{DM}.  
We have also introduced $N_i = N n_i$, where
$n_i$ is the index of the $i$th vertex of the Dynkin diagram.  Equivalently,
one may think of $i$ as running through the irreducible 
representations ${\bf r}_i$
of $\Gamma$, in which case $n_i$ can be thought of as the dimension
of  ${\bf r}_i$.
The dual Coxeter number of the corresponding ADE Lie group is
$h = \sum_i n_i$.
The number of elements of $\Gamma$ is $\ord(\Gamma) = \sum_i n_i^2$.

The field content of the gauge theory can be summarized conveniently with a 
quiver diagram which for the simple cases under consideration here is 
nothing but the corresponding extended Dynkin diagram.  
For each vertex in the Dynkin diagram, we have an ${\cal N}=2$
vector multiplet transforming under the adjoint of $U(N_i)$.  For each
line in the diagram, there is a bifundamental hypermultiplet $a_{ij}$ in
the representation $(N_i, \bar N_j)$.

To write a superpotential for this gauge theory, it is convenient to
decompose the fields into ${\cal N}=1$ multiplets.  Each $a_{ij}$ will
give rise to a pair of chiral multiplets, $(B_{ij}, B_{ji})$, where
$B_{ij}$ is a complex matrix transforming in the $(N_i, \bar N_j)$ 
representation.  Moreover, there is a chiral multiplet $\phi_i$ 
for each vector multiplet in the theory.

The superpotential is then
\be
W = \sum_{i} \Tr \mu_i \phi_i 
\label{suppot}
\ee
where $\mu_i$ is the ``complex moment map"
\be
{{\mu_i}^{\alpha_i}}_{\beta_i} = 
\sum_j s_{ij} {{B_{ij}}^{\alpha_i}}_{\gamma_j}
{{B_{ji}}^{\gamma_j}}_{\beta_i} \ .
\label{mmap}
\ee
Although the indices are confusing, essentially all we have done is construct
a cubic polynomial in the ${\cal N}=1$ superfields consistent with
${\cal N}=2$ SUSY and the gauge symmetry.  The factor $s_{ij}$ is the 
antisymmetric adjacency matrix for the Dynkin diagram: $s_{ij}=\pm 1$ when
$i$ and $j$ are adjacent nodes and zero otherwise.  The upper index $\alpha_i$
indicates a fundamental representation of $U(N_i)$, while a lower index
$\beta_i$ indicates an anti-fundamental representation of $U(N_i)$.
There is a relation among the $\mu_i$
\be
\sum_i \Tr \mu_i = 0 \ .
\label{consmu}
\ee 
which holds because the trace gives something
symmetric in $i$ and $j$ summed against $s_{ij}$ which is
antisymmetric.

This ${\cal N}=2$ gauge theory is superconformal and thus must have an
$R$ symmetry.
As a result, the superpotential term in the action 
$\int d^2\theta \, W$ will have $R$ charge zero. 
We take the convention that the spinor $\theta$ has $R$ charge $1$.  Therefore
$W$ must have $R$ charge $2$.  Conveniently, the $B_{ij}$ and the $\phi_i$ have
$R$ charge $2/3$ and the superpotential, as noted above, is cubic. 

In the large $N$ limit in the case of D3-branes, we can invoke the AdS/CFT 
correspondence for this gauge theory \cite{LNV, KS}.  The correspondence
tells us that the gauge theory described above is dual to type IIB
supergravity (SUGRA) on an $AdS_5 \times \S^5/\Gamma$ 
background.  To see how the
orbifolding works, consider 
$\S^5 = \{(z_1, z_2, z_3)\in \C{3}: \sum_i |z_i|^2=1\}$.
The group $\Gamma$ acts only on $(z_1, z_2) \in \C{2}$.  As a result, 
there is an $\S^1$ of the $\S^5$ which is left invariant under $\Gamma$.

We can add a term to the superpotential
(\ref{suppot}) that will give masses $m_i$ to the $\phi_i$.  Such a term
will eliminate the $\phi_i$ from the theory
at energies below the mass scale set by the $m_i$ and 
break the supersymmetry from ${\cal N}=2$ to ${\cal N}=1$.  In particular,
we add the term
\[
W' = W - \frac{1}{2} \sum_i m_i \Tr \phi_i^2 \ .
\]  
To see what happens at low energies, let us look at the equations of motion
$dW'=0$.  By varying with respect to the matter fields that define $\mu_i$,
we see that 
\be
\phi_i = \phi \Id_{N_i} 
\label{phii}
\ee
where $\phi$ is the Lagrange multiplier used to ensure
that the constraint (\ref{consmu}) is satisfied.
Varying with respect to $\phi_i$ and employing (\ref{phii}), one gets
\[
\mu_i = m_i \phi \Id_{N_i} \ .
\]
From (\ref{consmu}), it follows that
\[
\sum_i n_i m_i = 0 \ .
\]

Assuming that none of the $m_i=0$, we can eliminate $\phi_i$ from the action
to find an effective low energy superpotential:
\[
W_{eff} = \sum_i \frac{1}{2m_i} \Tr \mu_i^2 \ .
\]
Notice that $W_{eff}$ is quartic in the superfields $B_{ij}$ 
(see (\ref{mmap})).  
We would like the endpoint of the RG flow 
generated by adding
these mass terms to be an IR conformal fixed point.  Superconformal
theories preserve the $R$ symmetry.  However, if the fields $B_{ij}$ are given
their naive $R$ charges of $2/3$, this quartic superpotential will 
explicitly break our $R$ symmetry.  By giving the $B_{ij}$ anomalous dimensions,
we find that after flowing to the IR, the $R$ charge of the $B_{ij}$ can
be adjusted to $1/2$, and the $R$ symmetry is preserved.

In the context of AdS/CFT correspondence, the authors of 
\cite{GNS} generalized an argument of \cite{KW} for the $A_1$
case, arguing that the IR endpoint of this RG flow is dual to
type IIB SUGRA in an $AdS_5 \times \bX_\Gamma$ background 
where the $\bX_\Gamma$ are 
the level surfaces of certain ``generalized
conifolds''.  The generalized conifolds are three complex dimensional 
Calabi-Yau manifolds with a conical scaling symmetry.  The conifolds can
be described by a polynomial embedding relation 
$F_\Gamma=0$ in $\C{4}$.  To conform
with the notation of \cite{GNS}, we use the coordinates $(x,y,z,\phi)\in \C{4}$.
The polynomial $F_\Gamma$ is invariant under a $\C{*}$ action, the real part of
which is the conical scaling symmetry while the imaginary part corresponds to
an $R$ symmetry transformation in the dual gauge theory.  
$F_\Gamma$ transforms under this $\C{*}$ 
action with weight $h$, the dual Coxeter number.  The coordinate $\phi$ 
will transform with weight one, and the remaining coordinates $x$, $y$, and $z$
transform with weights\footnote{One should really separate out
the $A$ series into two series so that one does not have
fractional weights. The series for $A_{2k}$ would have weights
$(2,2k+1,2k+1,2k+2)$. The volume formula gives the same answer
when one plugs in the fractional weights, so, for conciseness of
notation, we will not separate the series.}
\be
\renewcommand{\arraystretch}{1.2}
\begin{array}{ccccc}
\Gamma & \alpha = [x] & \beta=[y] & h/2 = [z] & h \\
A_k & 1 & \frac{k+1}{2} & \frac{k+1}{2} & k+1 \\
D_k & 2 & k-2 & k-1 & 2(k-1) \\
E_6 & 3 & 4 & 6 & 12 \\
E_7 & 4 & 6 & 9 & 18 \\
E_8 & 6 & 10 & 15 & 30 
\end{array}
\label{weights}
\ee
We will not need the explicit form of the polynomials $F_\Gamma$ in what
follows; we refer the interested reader to \cite{GNS} for more details.
To make things more concrete, however, we give 
\be
F_{A_k} = \prod_{i=0}^{k} (x - \xi_i \phi) + y^2 + z^2 \ ,
\label{ak}
\ee
\be
F_{D_k} = \prod_{i=0}^{k-2} (x - \xi_i \phi^2) + 
                           t_0 \phi^k y + xy^2 + z^2 \ ,
\ee
where $\xi_i$ and $t_0$ are free constants transforming with
weight zero under the $\C{*}$ action.  


\subsection{The Central Charge}

Conformal field theories are characterized by a number 
called the central charge $c$ which appears in many 
correlation functions.  From the AdS/CFT dictionary,
for a conformal field theory dual to an $AdS_5 \times \bX_5$ background,
we know that $c \sim 1/\Vol(\bX_5)$ \cite{Kostas}.  
If we can calculate all the relevant
volumes and central charges independently, we can make a check of the
AdS/CFT correspondence.  In particular, in the UV we have the
the orbifolded theory, $\bX_5 = \S^5/\Gamma$,
and in the IR we have the generalized conifold theory, $\bX_5 = \bX_\Gamma$.
It ought to be true that
\[
\frac{c_{IR}}{c_{UV}} = \frac{1/\Vol(\bX_\Gamma)}{1/\Vol(\S^5/\Gamma)} \ .
\]

In the case $\Gamma=A_1$, this calculation was done in \cite{Gubser}.  
%
We now attempt to check this formula for arbitrary $\Gamma$.
The volume of $\S^5/\Gamma$ is straightforward to compute.  Indeed,
$\Vol(\S^5) = \pi^3$, and to find the volume of $\S^5/\Gamma$ we just
divide by the number of elements of $\Gamma$, which is nothing but 
$\sum_i n_i^2 = 2 \alpha \beta$ (see (\ref{weights})).
$\Vol(\bX_5)$ can be calculated from (\ref{voltot}), but, for suspense,
we will leave this step to the very end.  

First, let us show that
\be
\frac{c_{IR}}{c_{UV}} = \frac{27}{32}
\label{circuv}
\ee
independent of $\Gamma$. We begin by recalling some basic facts about
conformal field theories.  
First, we define the central charge $c$ and another anomaly
coefficient $a$ in terms of the one point function
of the stress energy tensor
\[
\langle T^\mu_\mu \rangle = -a E_4 - c I_4
\]
where $E_4$ and $I_4$ are scalars quadratic in and depending
only on the Riemann curvature.  
We will not try to calculate
$a$ and $c$ explicitly, but rather just the ratio
$c_{IR}/c_{UV}$.  Let $R_\mu$ be the $R$ symmetry current.  It was shown
in \cite{AFGJ} that
\be 
\langle (\partial_\mu R^\mu) T_{\alpha \beta} T_{\gamma \delta} \rangle
\sim (a-c) \sim \sum_\psi r(\psi) \ , 
\ee
\be
\langle (\partial_\mu R^\mu) R_\alpha R_\beta \rangle
\sim (5a-3c) \sim \sum_\psi r(\psi)^3 \ .
\ee
The sum is over all the fermions $\psi$ in the gauge theory, and $r(\psi)$ is
the $R$ charge.

To proceed, we need to classify all of the fermions in the gauge theories
of interest along with their $R$ charges.  As noted above, in the UV orbifold
theory the chiral superfields $B_{ij}$ and $\phi_i$ have $R$ charge
$2/3$.  As a result, the fermions in these chiral multiplets will
have $R$ charge $-1/3$. 
Thus, for each line in the extended Dynkin diagram, we have $2 N_i N_j$
fermions with $R$ charge $-1/3$, and for each vertex of the diagram, we have
$N_i^2$ fermions with $R$ charge $-1/3$.  There are also for each vertex
$N_i^2$ gluinos in the theory,
the superpartners of the gauge fields.  As the gauge fields are uncharged under
the R symmetry, the gluinos will have $R$ charge $1$.  We can now compute the 
sum over the $R$ charges in the UV:
\be
a_{UV}-c_{UV} \sim \sum_\psi r(\psi) = \frac{2N^2}{3} \left( \sum_i n_i^2 -
 \sum_{\langle ij \rangle} n_i n_j \right)
\ ,
\ee
\be
5a_{UV}-3c_{UV} \sim \sum_\psi r(\psi)^3 = \frac{26N^2}{27} \sum_i n_i^2 -
\frac{2N^2}{27} \sum_{\langle ij \rangle} n_i n_j \ .
\ee
The sum over $\langle ij \rangle$ is limited to nearest neighbor 
nodes of the Dynkin diagram, \ie, nodes connected by a line.
Typically, for conformal field theories with AdS duals, we expect that
$a=c$.  Indeed, it is a property of the simply laced Dynkin diagrams 
we are considering that 
\[
\sum_i n_i^2 =
 \sum_{\langle ij \rangle} n_i n_j \ .
\]
Thus, $5a_{UV} - 3c_{UV} = 2c_{UV} \sim \frac{8N^2}{9} \sum n_i^2$.

When we flow to the IR, the $\phi_i$ 
fields will get a mass and disappear from the theory.  As a result, the 
$N_i^2$ fermions with $R$ charge $-1/3$ will disappear from the sum.  Above,
we noted that the $R$ charge of the superfields $B_{ij}$ changes to $1/2$.
Thus the $2N_i N_j$ fermions for each line in the Dynkin diagram will now
have $R$ charge $-1/2$.  The $R$ charge of the gluinos is unchanged.
Repeating the above calculation in these slightly different circumstances,
we find that 
\be
a_{IR}-c_{IR} \sim \sum_\psi r(\psi) =  N^2 \left( \sum_i n_i^2 -
 \sum_{\langle ij \rangle} n_i n_j \right) \ ,
\ee
\be
5a_{IR}-3c_{IR} \sim \sum_\psi r(\psi)^3 = N^2 \sum_i n_i^2 -
\frac{N^2}{4} \sum_{\langle ij \rangle} n_i n_j \ .
\ee
Thus, $a_{IR} = c_{IR}$ as expected, and, moreover, 
$5a_{IR} - 3c_{IR} = 2c_{IR} \sim \frac{3N^2}{4} \sum n_i^2$.
Dividing our two results yields (\ref{circuv}).

At this point, one can say that it is 
a prediction of the AdS/CFT correspondence that
\be
\Vol(\bX_\Gamma) = \frac{\Vol(T^{1,1}) \ord(A_1)}{\ord(\Gamma)} =
\frac{16 \pi^3}{27 \alpha \beta} \ .
\label{predict}
\ee
As the products of weights for all these conifolds is $\alpha\beta h/2$
and the index is always $2$, we see that our formula
(\ref{voltot}) gives precisely the correct answer.

\section{A Somewhat Strange Conjecture}

\subsection{Motivation}

In attempting to prove the volume formula
(\ref{voltot}), an interesting fact about the holomorphic monomials
on an affine cone cut out by a weighted homogeneous polynomial
was noticed.
The initial observation that began the set of ideas which follows was 
a recollection of Weyl's Law for the growth of the eigenvalues of the
Laplacian.  Let $\nabla^2 f = E f$, where $\nabla^2$ is the Laplacian on some
compact $D$ dimensional manifold $M_D$ and $E$ the corresponding 
energy or eigenvalue.  Weyl's Law states that the number of eigenfunctions 
with energy less than $E$ scales as
\[
N(E) \sim \gamma \Vol(M_D) E^{D/2} 
\]
for large $E$ where $\gamma$ is some constant of proportionality.  

Of course, without a metric, we can no sooner derive
an expression for $\nabla^2$ than calculate the volume.  However,
in the context of AdS/CFT correspondence, it is not the eigenfunctions
of the Laplacian, or equivalently the harmonic functions
on the cone over $\bX$, which play the most important role.  Most of these
harmonic functions have unprotected dimensions and energies which can change
when the coupling constant in the gauge theory changes.  
Instead it is the chiral primary operators (CPOs) which are the most important.
The CPOs have the maximum possible $R$ charge for a given dimension, and
the SUSY algebra protects their dimension when the coupling constant is changed.

As was noted above, the $U(1)$ $R$ symmetry corresponds to the imaginary part
of the $\C{*}$ action on $\bX$.  Let us consider the 
spaces $\bX_\Gamma$.  The dimension of an operator such as
$x^m \bar x^{\bar m} y^n \bar y^{\bar n} z^p z^{\bar p} \phi^q \bar \phi^{\bar q}$ 
is just 
\[
L = \alpha(m + \bar m) + \beta(n+\bar n) + \frac{h}{2}(p + \bar p) + q + \bar q \ .
\]
On the other hand, the $R$ charge is
\[
r = \alpha(m- \bar m) + \beta(n- \bar n) + \frac{h}{2}(p- \bar p) + q- \bar q \ .
\]
Clearly, the operators of this type that maximize the magnitude of the
$R$ charge are the purely
holomorphic or purely antiholomorphic monomials.  Let $\cP^F_L$ be the space
of holomorphic degree $L$ monomials quotiented by the relation $F =0$
where $F$ is the defining polynomial of the cone.
In the spirit of the Weyl scaling law, we conjecture that for large $L$, 
\be
\label{theconj}
\dim(\cP^F_L) \sim \gamma L^{n-1} \Vol(\bX)
\ee
where $\gamma$ is a constant of proportionality.\footnote{
We would like to thank Steve Gubser suggesting this possibility.
}

To calculate $\dim(\cP^F_L)$, note that 
\be
\label{dform}
\dim(\cP^F_L) = \dim(\cP_L) - \dim(\cP_{L-d})
\ee
where $\dim(\cP_L)$ is the number of holomorphic polynomials of degree
$L$ on $\C{n}$ and $d$ is the degree of $F$.  To leading order, 
$$
\dim(\cP_L) = \frac{d \cP_{<L}}{dL}
$$
where $\cP_{<L}$ is the number of polynomials with degree less than or 
equal to $L$.  Also to leading order, $\cP_{<L}$ is
the volume of a $n+1$ dimensional
pyramid with apex at the origin and legs along the positive
axes of $\R{n+1}$.  Each leg will have length $L/w_i$.
Thus
$$
\dim(\cP_L) = \frac{L^n}{n! w}
$$
where, as before, $w$ is the product of the weights.

Now, we apply (\ref{dform}) to obtain
\be
\dim(\cP_L^F) = L^{n-1} \frac{d}{(n-1)! w} + {\cal O}(L^{n-2}) \ .
\label{polyrel}
\ee
Comparing with (\ref{voltot}), we find that 
\be
\gamma = \frac{1}{2} \left(\frac{n }{\pi (|\bw| - d)} \right)^{n} \ .
\ee
Thus, the constant depends only on the index and the dimension of the 
Einstein space.

\subsection{Why does it Work?}

As has often been said, there are no coincidences in mathematics.
Thus, we would like to understand a deeper reason why the
conjecture turns out to be correct. We will see that, at least in
the smooth case, it is a consequence of the
Hirzebruch-Riemann-Roch (HRR) theorem. We will not attempt to
generalize to the orbifold case, except to note that the
HRR theorem has been generalized by Kawasaki to
orbifolds in \cite{Kaw}.

In order to proceed, we will first express the dimension of
$\cP^{F}_{L}$ in an algebraic geometric form. Recall
that there exists a map
$$
\mathrm{Sym}^{L}({\C{n+1}}^{*}) \longrightarrow H^{0}(\CP{n},\mO(H^{L})).
$$
where $H$ is the hyperplane bundle over the projective space.
Thus, we can compute the dimension of $\cP_{L}$ by computing the
dimension of the space of sections of a line bundle over $\CP{n}$. All
that is left to do is to impose the relation $F(\bz) = 0$. However,
$F$ defines a section of $H^{d}$, and its zero locus defines the
variety $\bV$ in $\CP{n}$. We can impose the relation by
restricting the line bundle $H^{L}$ to the variety $\bV$ and
computing its dimension:
\be
\label{dimform}
\dim(\cP^{F}_{L}) = H^{0}(\bV,\mO(H^{L})).
\ee
A useful reference for the algebraic geometry in this section is
\cite{GH}. A nice introduction to the Atiyah-Singer
index theorem and its application to the Dolbeault complex is
contained in \cite{Nash}.

We will now convert (\ref{dimform}) into an integral formula through
the use of the HRR theorem. We first recall the definition of the
Euler character of the twisted Dolbeault complex on a manifold
$M$ with line bundle $\mcL$:
\be
\label{echar}
\chi \equiv \sum_{q} (-1)^{q} \dim\ H^{0,q}_{\bar\del}(M,\mcL)
= \sum_{q} (-1)^{q} h^{0,q}(\mcL) \ .
\ee
Here, the forms take values in the line bundle $\mcL$. 
The HRR theorem states that
$$
\chi = \int_{M} ch(\mcL) td(M)
$$
where $ch$ and $td$ are characteristic classes to be defined later.

Because we are on a K\"ahler manifold, the twisted Hodge numbers obey
$h^{p,q}(\mcL) = h^{q,p}(\mcL)$, and we can interchange the indices in
(\ref{echar}).
The next thing we need is the Dolbeault theorem
which relates the Dolbeault cohomology groups of a manifold to certain
sheaf cohomology groups:
$$
H^{p,q}_{\bar\del}(M,\mcL) \cong H^{q}(M,\Omega^{p}(\mcL)).
$$
$\Omega^{p}(\mcL)$ denotes the sheaf of holomorphic $p$-forms on $M$
valued in $\mcL$.
Putting all this together, we obtain the identity:
$$
\chi = \chi(\mO(\mcL)) \equiv \sum_{q} (-1)^{q} \dim\
H^{q}(M,\mO(\mcL)).
$$

Next, we recall another fact from algebraic geometry, the Kodaira
vanishing theorem.
This states that, for any positive line bundle $\mcL$, not
necessarily the same $\mcL$ as before,
over a manifold of complex dimension $m$,
$$
H^{q}(M,\Omega^{p}(\mcL)) = 0 \quad \mathrm{when}\quad p+q>m.
$$
Specialize to $M=\bV$ and let $p=n-1$. Recall that
$\Omega^{n-1} = K_{\bV}$, the canonical bundle on $\bV$. Then, we have
$$
H^{q}(\bV,\mO(K_{\bV} \otimes \mcL)) = 0 \quad \mathrm{for} \quad q > 0.
$$
Let $\mcL = K_{M}^{-1} \otimes H^{L}$. This is positive
because $K_{M}^{-1}$ and the hyperplane bundle $H$ are both positive.
Hence, all $H^q(\bV,\mO(H^L))$ vanish for $q>0$, and
$$
\chi(\mO(H^{L})) = \dim\ H^{0}(M,\mO(H^{L})).
$$
This gives us our integral,
$$
\dim(\cP^{F}_{L}) = \dim\ H^{0}(\bV,\mO(H^{L})) = \chi(\mO(H^{L}))
= \int_{\bV} ch(H^{L}) td(\bV).
$$

The Chern character and Todd class
are given by infinite series that begin as follows (see, for
example \cite{Nak,Nash} and note that all higher Chern classes of
$H$ vanish as it is a line bundle):
\bear
ch(H) &=& 1 + c_{1}(H) + \frac{1}{2}c_{1}(H)^{2} + \cdots \nn\\ 
td(\bV) &=& 1 + \frac{1}{2}c_{1}(\bV) + \frac{1}{12}(c_{1}(\bV)^{2} +
c_{2}(\bV)) + \cdots \nn
\eear
From the adjunction formula we have
$$
c_{1}(H) = \frac{1}{n+1-d} c_{1}(\bV) \ .
$$
Thus, everything can be expressed as integrals over various Chern 
classes of $\bV$. The key point to notice is that the series for
$ch(H)$ contains only first Chern classes of $\bV$. Thus, when we
multiply out the two series and take the $2(n-1)$th degree form, the
answer will be of the form:
\be
\label{almost}
\dim(\cP^{F}_{L}) = Q(L) \int c_1(V)^{n-1} + \mcO(L^{n-2}) \ 
\ee
where $Q(L)$ is some $(n-1)$th degree polynomial in $L$ with rational
coefficients. It is straightforward to compute the leading term in
$L$ as it only comes from the expression for the Chern character.
If we write $c_1(H) = x$, then, by definition, $ch(H) = e^x$.
Therefore, $ch(H^L) = e^{Lx}$ and the $n$th order term is $L^n x^n /
n!$. As $x = c_1(V)/(n+1-d)$, we can see that the leading term in $Q(L)$
must be $L^{n-1} / ((n-1)!(n+1-d)^{n-1})$.

Now, recall formula (\ref{volalm}) for the volume of $\bV$ in terms
of the integral of Chern classes. We multiply this by the result for
the length of the fiber to obtain
\be
\label{volch}
\Vol(\bX) = \frac{2(n+1-d)}{(n-1)!} \left(\frac{\pi}{n} \right)^{n} 
            \int c_1(\bV)^{n-1} \ .
\ee
Combining (\ref{volch}) with (\ref{almost}), we get
\be
\label{conjform}
\dim(\cP^{F}_{L}) \sim \frac{L^{n-1}}{2} \left(\frac{n}
{\pi(n+1-d)}\right)^{n} \Vol(\bX)
\ee
for large $L$.
Thus, we have explained the observation (\ref{polyrel}) for all 
Einstein manifolds constructed from $F(\bz)=0$ where $F(\bz)$ is
a homogeneous polynomial. 
In the weighted case, we expect $n+1-d$ would
be replaced by the index $|\bw|-d$.
A similar explanation involving the generalized HRR theorem ought to hold 
for these more general examples.

\section{Discussion}

We have seen that given remarkably little knowledge about a cone,
we can compute the volume of the base.
If the cone is smooth except at the tip and admits a
Calabi-Yau metric, we can usually compute the volume in terms of 
the dimension, the degree of the defining polynomial
and the index, $|\bw|-d$.
This statement is impressive considering how many cones
of a given degree, index and dimension exist.
For example, the $A_k$ series from (\ref{ak}) in general also depends 
on $k+1$ variables, $\xi_i$, 
but the volume is independent of these deformations. Essentially,
we have shown that the volume can be determined almost solely in
terms of topological numbers of the manifold.
We have also seen that CPOs have an intriguing
relationship to sections of a line bundle. It would be
interesting to explore this relationship further and to see what
additional elements of the gauge theory can be related to the
topology of the base.


Given recent results about the existence of
K\"ahler-Einstein metrics on various weighted projective
varieties \cite{JK1,JK2} and Sasaki-Einstein manifolds over them
\cite{BG3,BG4}, there exists a vast new array of spaces on
which to examine the AdS/CFT correspondence, extending the list
of \cite{MoPle}. It would be very interesting to see if the
techniques such as those used in this
paper can be extended to give more information about the correspondence
for these cones. For example, understanding in precise and
quantitative detail the homology
structure of the base would allow one to investigate fractional
branes on these spaces.

\section*{Acknowledgments}
C.~P.~H. would like to thank Steve Gubser for collaboration in
the early stages of this project.  Many thanks go also to
Igor Klebanov for discussions and help with the manuscript.
We would like to thank Chris Beasley, Brent Doran, J\'anos Koll\'ar
and John Pearson for many useful discussions. Figure \ref{fig1}
is used with permission from \cite{GNS}. This work was supported
under NSF grant PHY-9802484.  C.~P.~H. was supported in part
by the Department of Defense.

\appendix

\section{The Integral of Chern Classes}

We will now compute the integral of Chern classes (\ref{chint}).
We would like to thank J\'anos Koll\'ar for giving us this
argument. Any mistakes are our own.

Recall that the integral of a product of Chern classes of
ordinary line bundles is simply
the intersection numbers of the varieties defined by their zero
loci. Thus, we would like to convert the integral into something
we can compute in terms of intersections. If $I = |\bw| - d$,
then we are dealing with V-bundles that are the $I$th power of
the hyperplane V-bundle. The Chern classes in our integral are
defined in terms of the curvature of a connection on the bundle
that lives on the uniformizing chart of the orbifold.
Thus, following the usual arguments, it easy to see that, for any
line V-bundle $\mcL$, $c_1(\mcL^n) = n c_1(\mcL)$. This gives
$$
\int_\bV c_1(\bV)^{n-1} = I^{n-1} \int_\bV c_1(H)^{n-1}.
$$
We can convert this into an integral over the entire space by
multiplying by the Poincar\'e dual of the hypersurface, $c_1(H^d)
= d c_1(H)$, giving
$$
\int_\bV c_1(\bV)^{n-1} = d I^{N-1} \int_\WP{\bw} c_1(H)^n.
$$

Unfortunately, $H$ is still not a line bundle. However, its $w$th
power (recall that $w$ is the product of the weights) is one as all
the weights divide $w$. Thus, we can compute
$$
\int_\WP{\bw} c_1(H)^n = \frac{1}{w^n} \int_\WP{\bw} c_1(H^w)^n.
$$
Now, we can relate this to the intersection number of a generic
section of $H^w$. This is simple enough to compute explicitly. Let us
examine the following sections of the line bundle:
$$
z_i^{w/w_i} - z_0^{w/w_0}
$$
where the $z_i$ are the weighted homogeneous coordinates on the weighted
projective space. To determine the zero locus of this section, we
first use the $\C{*}$ action to set $z_0 = 1$. Then, we see that
the solutions are given by $z_i = \z_{w/w_i,a}$ where $\z_{n,a}$,
$a\in 0\ldots n-1$,
are the $n$th roots of unity. Thus, the points on the common
intersection of the zero locus of all these sections are given by
$$
(1,\z_{w/w_1,a_1},\ldots,\z_{w/w_n,a_n}).
$$
As there are $w/w_i$ possibilities for each $a_i$, this gives
$w_0 w^{n-1}$ solutions. However, we have overcounted because
this choice of coordinates is not unique. In fact, we can
act with $\z_{w_0,a_0}$ under the weighted $\C{*}$ action and keep a
one in the first position. Thus, we divide the number of solutions
by the factor $w_0$
giving the total number of intersection points as $w^{n-1}$.
Finally, we put this all together and obtain
$$
\int_\bV c_{1}(\bV)^{n-1} = \frac{d}{w} I^{n-1}
$$
which is what we sought to prove.

One can also see this result by looking at the projective cover of
weighted projective space. Given our set of weights, we define
the following new set
of variables: $t_i = z_i^{1/w_i}$. Under the weighted
action, the $t_i$ transform uniformly and, as such, are
coordinates on the normal projective space $\CP{n}$. However,
the map we have defined here is not 1-1. As we
circle the origin in the plane of one of the $t_i$, we circle
the origin in
the $z_i$ plane $w_i$ times. Thus, in order to obtain the
weighted projective space, we have to quotient by the group
$\Z_{w_0} \times \cdots \times \Z_{w_n}$. The order of this
group is simply
the product of the weights, $w$. However, the degree of the
cover is
this value divided by the greatest common divisor of the
weights because that is the order of the element $(1,\cdots,1)$
in the group. In all the situations we will deal with, the gcd
will be 1 as, otherwise, the weighted projective space would not
be well-formed.

If we now look at the inverse image of the hypersurface, it is a
degree $d$ hypersurface in an ordinary projective space.
Unfortunately the metric is no longer K\"ahler-Einstein.
However, we can still write the volume as the integral of the
$n$th power of a differential form. If one accepts that this is
$I [\varpi]$ in cohomology where $\varpi$ is the pullback of the
K\"ahler form of the Fubini-Study metric to the projective
variety, then we can compute the volume of the
covering hypersurface. Then, we divide by the degree of the
cover, $w$, to give the same result as above. 

\section{The Volume of $V_{5,2}$}

We begin by deriving a Ricci-flat metric on the cone over $V_{5,2}$:
\begin{equation}
\sum_{i=1}^{5} w_i^2 = 0 \ .
\label{conifold}
\end{equation}
The $w_i$ are assumed to be complex variables.  A reason we can
even hope to find an explicit metric is that $V_{5,2}$ is a coset 
manifold, $SO(5)/SO(3)$.  We proceed by generalizing an argument
contained in \cite{KW}.  Note, metrics on $V_{5,2}$ have 
appeared in \cite{CGLP, Stenzel}.

As (\ref{conifold}) is symmetric under $SO(5)$ rotations of the 
five complex variables, we assume the K\"ahler potential $K$ also possesses this
$SO(5)$ symmetry.  
There is only one $SO(5)$ invariant length in the problem, 
\[
\rho = \sum_{i=1}^{5} |w_i|^2,
\]
and $K = f(\rho)$ can be a function only of $\rho$.  Indeed
as we will now see, we can further
assume that $K = \rho^a /2$ because $K$ must transform homogeneously
under the scaling $z_i \to \lambda z_i$ where $\lambda \in \C{*}$.
The factor of $1/2$ is added for later convenience.

On a 
Calabi-Yau manifold of complex dimension 4, there exist nonvanishing 
holomorphic and antiholomorphic 4-forms whose wedge product is proportional 
to the volume form on the manifold.  For our 4-fold, the holomorphic
4-form is
\[
\Omega = \frac{dw_1 \wedge dw_2 \wedge dw_3 \wedge dw_4}{w_5}.
\]
On the other hand, the volume form may also be computed from the 
wedge product of the K\"ahler form $\omega = \partial \bar \partial K$:
\[
\omega \wedge \omega \wedge \omega \wedge \omega \sim \Omega \wedge \bar\Omega \ .
\]
Counting powers, we find that $a = 3/4$.  
This gives us the metric on $V_{5,2}$:
$$
g_{i\bar{\jj}} = \partial_i \bar{\partial}_j \rho^{3/4} / 2.
$$

The next step is to rewrite the metric in a way that makes 
the volume calculation easier.
In order to write the metric in conical form
\[
ds^2 = dr^2 + r^2 ds_{V_{5,2}}^2 \ ,
\]
a scaling argument necessitates that 
we define the radius of the cone as $r^2=\rho^{3/4}$.  
Inspired by the change of variables needed
to write the conifold as a cone over $T^{1,1}$, we have found
a change of variables that isolates the angular
part of the metric.  In particular
\begin{eqnarray*}
w_1 &=& \frac{1}{\sqrt{2}} r^{4/3} e^{i\psi}
\left( \Lambda_+ 
\cos \left( \frac{\theta_1 + \theta_2}{2} \right)
\cos \left( \frac{\phi_1 + \phi_2}{2} \right)
+ \Lambda_- 
\cos \left( \frac{\theta_1 - \theta_2}{2} \right)
\sin \left( \frac{\phi_1 + \phi_2}{2} \right)
\right) \\
w_2 &=& \frac{1}{\sqrt{2}} r^{4/3} e^{i\psi}
\left( -\Lambda_+ 
\cos \left( \frac{\theta_1 + \theta_2}{2} \right)
\sin \left( \frac{\phi_1 + \phi_2}{2} \right)
+ \Lambda_- 
\cos \left( \frac{\theta_1 - \theta_2}{2} \right)
\cos \left( \frac{\phi_1 + \phi_2}{2} \right)
\right) \\
w_3 &=& \frac{1}{\sqrt{2}} r^{4/3} e^{i\psi}
\left( -\Lambda_+ 
\sin \left( \frac{\theta_1 + \theta_2}{2} \right)
\cos \left( \frac{\phi_1 - \phi_2}{2} \right)
+ \Lambda_- 
\sin \left( \frac{\theta_1 - \theta_2}{2} \right)
\sin \left( \frac{\phi_1 - \phi_2}{2} \right)
\right) \\
w_4 &=& \frac{1}{\sqrt{2}} r^{4/3} e^{i\psi}
\left( -\Lambda_+ 
\sin \left( \frac{\theta_1 + \theta_2}{2} \right)
\sin \left( \frac{\phi_1 - \phi_2}{2} \right)
- \Lambda_- 
\sin \left( \frac{\theta_1 - \theta_2}{2} \right)
\cos \left( \frac{\phi_1 - \phi_2}{2} \right)
\right) \\
w_5 &=& -\frac{1}{\sqrt{2}} r^{4/3} e^{i\psi} \sin(\alpha)
\end{eqnarray*} 
where
\[
\Lambda_\pm =
\cos \alpha \cos \left(\frac{\beta}{2}\right) \pm 
i \sin \left(\frac{\beta}{2}\right).
\]
At the locus of points $\alpha=0$ and $\beta=0$, we recover the
standard Euler angle parameterization on $T^{1,1}$.  The allowed
ranges of the new set of variables are 
\[
0<r \;,\;\; 0 \leq \theta_i < \pi \;,\;\; 0 \leq \phi_i < 2\pi
\;,\;\;
0 \leq \psi < 2\pi
\;,\;\;
0 \leq \alpha < \pi/2
\;,\;\;
0 \leq \beta < 4\pi
\]
Define the one forms:
\begin{eqnarray*}
e^{\psi} &=& (d\psi + \frac{1}{2}\cos\alpha (d\beta - \cos\theta_1 d\phi_1 -
\cos\theta_2 d\phi_2)) \\
e^{\beta} &=& (d\beta - \cos\theta_1 d\phi_1 - \cos\theta_2 d\phi_2) \\
e^{\phi_i} &=& \sin\theta_i d\phi_i
\end{eqnarray*}
Letting Maple handle the dirty work, we find that
the metric on $V_{5,2}$ may be written in angular coordinates as
\begin{eqnarray*}
ds_{V_{5,2}}^2 &=&
 \frac{9}{16} (e^{\psi})^2 + \frac{3}{8} d\alpha^2
+ \frac{3}{32} \sin^2 \alpha (e^{\beta})^2 \\
& & + \frac{3}{32}(1+\cos^2 \alpha)((e^{\phi_1})^2 + (e^{\phi_2})^2 +
d\theta_1^2 + d\theta_2^2)
+ \frac{3}{16} \sin^2 \alpha \cos\beta e^{\phi_1}e^{\phi_2} \\
& &- \frac{3}{16} \sin^2 \alpha \cos\beta d\theta_1 d\theta_2 
+ \frac{3}{16} \sin^2 \alpha \sin\beta (d\theta_1 e^{\phi_2} +
d\theta_2 e^{\phi_1}).  
\end{eqnarray*} 
The determinant of this metric is
\[
\sqrt{\mbox{det} g} = \frac{3^4}{2^{14}} \sin \alpha \cos^2 \alpha 
\sin \theta_1 \sin \theta_2.
\]
As this metric is the base of a Calabi-Yau cone, we expect it to be 
Einstein.  In fact, $R_{ij} = 6g_{ij}$.  Integrating the volume 
form, we find that the volume of $V_{5,2}$ is 
\[
\Vol(V_{5,2}) = \frac{27}{128} \pi^4.
\] 

\end{document}